\documentclass[prd,nofootinbib,showpacs,preprint]{revtex4}
\usepackage{amsmath,amssymb}
\usepackage{epsfig}
\usepackage{graphicx}
\usepackage[usenames,dvipsnames]{color}
\usepackage{slashed}
\usepackage{datetime}
\usepackage{amsmath}
\usepackage{pdfpages}
\usepackage{color}

\begin{document}
\title{Scalar Dark Matter with Type II Seesaw}
\author{Arnab Dasgupta}
\email{arnab@ctp-jamia.res.in}
\affiliation{Centre for Theoretical Physics, Jamia Millia Islamia - Central University, Jamia Nagar, New Delhi - 110025, India}
\author{Debasish Borah}
\email{dborah@tezu.ernet.in}
\affiliation{Department of Physics, Tezpur University, Tezpur - 784028, India}

\begin{abstract}
We study the possibility of generating tiny neutrino mass through a combination of type I and type II seesaw mechanism within the framework of an abelian extension of standard model. The model also provides a naturally stable dark matter candidate in terms of the lightest neutral component of a scalar doublet. We compute the relic abundance of such a dark matter candidate and also point out how the strength of type II seesaw term can affect the relic abundance of dark matter. Such a model which connects neutrino mass and dark matter abundance has the potential of being verified or ruled out in the ongoing neutrino, dark matter as well as accelerator experiments.
\end{abstract}
\maketitle
\section{Introduction}
Recent discovery of the Higgs boson at the large hadron collider (LHC) experiment has established the standard model (SM) of particle physics as the most successful fundamental theory of nature. However, despite its phenomenological success, the SM fails to address many theoretical questions as well as observed phenomena. Three most important observed phenomena which the SM fails to explain are neutrino oscillations, matter-antimatter asymmetry and dark matter. Neutrino oscillation experiments in the last few years have provided convincing evidence in support of non-zero yet tiny neutrino masses \cite{PDG}. Recent neutrino oscillation experiments T2K \cite{T2K}, Double ChooZ \cite{chooz}, Daya-Bay \cite{daya} and RENO \cite{reno} have not only made the earlier predictions for neutrino parameters more precise, but also predicted non-zero value of the reactor mixing angle $\theta_{13}$. Matter-antimatter asymmetry of the Universe is encoded in the baryon to photon ratio measured by dedicated cosmology experiments like Wilkinson Mass Anisotropy Probe (WMAP), Planck etc. The latest data available from Planck mission constrain the baryon to photon ratio \cite{Planck13} as
\begin{equation}
Y_B \simeq (6.065 \pm 0.090) \times 10^{-10}
\label{barasym}
\end{equation} 
Presence of dark matter in the Universe is very well established by astrophysics and cosmology experiments although the particle nature of dark matter in yet unknown. According to the Planck 2013 experimental data \cite{Planck13}, $26.8\%$ of the energy density of the present Universe is composed of dark matter. The present abundance or relic density of dark matter is represented as
\begin{equation}
\Omega_{\text{DM}} h^2 = 0.1187 \pm 0.0017
\label{dm_relic}
\end{equation}
where $\Omega$ is the density parameter and $h = \text{(Hubble Parameter)}/100$ is a parameter of order unity.

Several interesting beyond standard model (BSM) frameworks have been proposed in the last few decades to explain each of these three observed phenomena in a natural way. Tiny neutrino masses can be explained by seesaw mechanisms which broadly fall into three types : type I \cite{ti}, type II \cite{tii} and type III \cite{tiii}. Baryon asymmetry can be produced through the mechanism of leptogenesis which generates an asymmetry in the leptonic sector first and later converting it into baryon asymmetry through electroweak sphaleron transitions \cite{sphaleron}. The out of equilibrium CP violating decay of heavy Majorana neutrinos provides a natural way to create the required lepton asymmetry \cite{fukuyana}. There are however, other interesting ways to create baryon asymmetry: electroweak baryogenesis \cite{Anderson:1991zb}, for example. The most well motivated and widely discussed particle dark matter (for a review, please see \cite{Jungman:1995df}) is the weakly interacting massive particle (WIMP) which interacts through weak and gravitational interactions and has mass typically around the electroweak scale. Weak interactions kept them in equilibrium with the hot plasma in the early Universe which at some point of time, becomes weaker than the expansion rate of the Universe leading to decoupling (or freeze-out) of WIMP. WIMP's typically decouple when they are non-relativistic and hence known as the favorite cold dark matter (CDM) candidate.

Although the three observed phenomena discussed above could have completely different particle physics origin, it will be more interesting if they have a common origin or could be explained within the same particle physics model. Here we propose a model which has all the ingredients to explain these three observed phenomena naturally. We propose an abelian extension of SM (for a review of such models, please see \cite{langacker}) with a gauged $B-L$ symmetry. Neutrino mass can be explained by both type I and type II seesaw mechanisms. Some recent works related to the combination of type I and type II seesaw can be found in \cite{typeI+II}. Dark matter can be explained due to the existence of an additional Higgs doublet, naturally stable due to the choice of gauge charges under $U(1)_{B-L}$ symmetry. Unlike the conventional scalar doublet dark matter models, here we show how the origin neutrino mass can affect the dark matter phenomenology. Some recent works motivated by this idea of connecting neutrino mass and dark matter can be found in \cite{nuDM}. In supersymmetric frameworks, such scalar dark matter have been studied in terms of sneutrino dark matter in type I seesaw models \cite{susy1} as well as inverse seesaw models \cite{susy2}. We show that in our model, the dark matter abundance can be significantly altered due to the existence of a neutral scalar with mass slightly larger than the mass of dark matter, allowing the possibility of coannihilation. And interestingly, this mass splitting is found to be governed by the strength of type II seesaw term of neutrino mass in our model. We show that for sub-dominant type II seesaw term, dark matter relic abundance can get significantly affected due to coannihilation whereas for dominant type II seesaw, usual calculation of dark matter relic abundance follows incorporating self-annihilation of dark matter only.

This paper is organized as follows: in section \ref{model}, we outline our model with particle content and relevant interactions. In section \ref{numass}, we briefly discuss the origin of neutrino mass in our model. In section \ref{darkmatter}, we discuss the method of calculating dark matter relic abundance. In section \ref{results}, we discuss our results and finally conclude in section \ref{conclude}.

\section{The Model}
\label{model}
We propose a $U(1)_{B-L}$ extension of the standard model with the particle content shown in table \ref{table1}. Apart from the standard model fermions, three right handed neutrinos $\nu_R$ are added with lepton number $1$. This is in fact necessary to cancel the $U(1)_{B-L}$ anomalies. Among the scalars, $H$ is the standard model like Higgs responsible for giving mass to fermions and breaking electroweak gauge symmetry. The second Higgs doublet $\phi$ does not acquire vacuum expectation value (vev) and also has no coupling with the fermions. This will act like an inert doublet dark matter in our model whose stability is naturally guaranteed by the gauge symmetry. The scalar triplet $\Delta$ serves two purposes: its neutral component contributes to the light neutrino masses by acquiring a tiny $(\sim \text{eV})$ vev and also generates a mass splitting between the CP-even and CP-odd neutral scalars in the inert Higgs doublet $\phi$.
\begin{center}
\begin{table}
\caption{Particle Content of the Model}
\begin{tabular}{|c|c|c|}
\hline
Particle & $SU(3)_c \times SU(2)_L \times U(1)_Y$ & $U(1)_{B-L}$ \\
\hline
$ (u,d)_L $ & $(3,2,\frac{1}{3})$ & $\frac{1}{3}$ \\
$ u_R $ & $(\bar{3},1,\frac{4}{3})$ & $\frac{1}{3}$ \\
$ d_R $ & $(\bar{3},1,-\frac{2}{3})$ & $\frac{1}{3}$ \\
$ (\nu, e)_L $ & $(1,2,-1)$ & $-1$  \\
$e_R$ & $(1,1,-2)$ & $-1$ \\
$\nu_R$ & $(1,1,0)$ & $-1$ \\
\hline
$H$ & $(1,2,1)$ & $0$ \\
$\phi$ & $(1,2,1)$ & $1$  \\
$ \Delta$ & $(1,3,2)$ & $2$  \\
$ S$ & $(1,1,0)$ & $2$  \\
\hline
\end{tabular}
\label{table1}
\end{table}
\end{center}
The Yukawa Lagrangian for the above particle content can be written as
\begin{eqnarray*}
\mathcal{L}_Y &=& Y_e \bar{L} H e_R + Y_{\nu}\bar{L}H^{\dagger}\nu_R+Y_d \bar{Q}Hd_R +Y_u \bar{Q}H^{\dagger}u_R \\
&& +Y_R S \nu_R \nu_R + f \Delta L L 
\label{yuklag}
\end{eqnarray*}
The gauge symmetry of the model does not allow any coupling of the inert Higgs doublet $\phi$ with the fermions. The scalar Lagrangian of the model can be written as
\begin{eqnarray*}
\mathcal{L}_H &=& \frac{\lambda}{4} \left ( H^{\dagger i}H_i - \frac{v^2}{2} \right )^2 +m^2_1 (\phi^{\dagger i}\phi_i ) + \lambda_{\phi} (\phi^{\dagger i}\phi_i )^2 \\
&& + \lambda_1 (H^{\dagger i}H_i)(\phi^{\dagger j}\phi_j ) + \lambda_2 (H^{\dagger i}H_j)(\phi^{\dagger j}\phi_i )+ \mu_{\phi \Delta} (\phi \phi \Delta^{\dagger} + \phi^{\dagger} \phi^{\dagger} \Delta ) \\
&& + \lambda_3 (H H \Delta^{\dagger} S + H^{\dagger}H^{\dagger} \Delta S^{\dagger} ) + m^2_2 S^{\dagger}S + \lambda_S (S^{\dagger} S)^2 + m^2_{\Delta} \Delta^{\dagger}\Delta +\lambda_{\Delta} (\Delta^{\dagger}\Delta)^2\\
&& +\lambda_4 (H^{\dagger}H)(S^{\dagger}S)+\lambda_5 (\phi^{\dagger}\phi )( S^{\dagger}S)+ \lambda_6 (\Delta^{\dagger}\Delta )(S^{\dagger}S)+\lambda_7 (H^{\dagger}H)(\Delta^{\dagger}\Delta )+\lambda_8 (\phi^{\dagger}\phi )(\Delta^{\dagger}\Delta )
\end{eqnarray*}
Assuming that the inert doublet $\phi$ does not acquire any vev, the neutral scalar masses corresponding to the Higgs doublets $H, \phi$ can be written as
$$ m^2_h = \frac{1}{2}\lambda v^2$$
$$m^2_{H_0} = m^2_1 +\frac{1}{2}(\lambda_1+\lambda_2)v^2+2\mu_{\phi \Delta} v_L$$
$$ m^2_{A_0} = m^2_1+\frac{1}{2}(\lambda_1+\lambda_2)v^2-2\mu_{\phi \Delta} v_L$$
where $m_h$ is the mass of SM like Higgs boson which is approximately 126 GeV and $v$ is the vev of the neutral component of SM like Higgs doublet $H$. The CP-even $(H_0)$ and CP-odd $(A_0)$ neutral components of inert doublet $\phi$ have a mass squared splitting proportional to $4\mu_{\phi \Delta} v_L$ where $v_L$ is the vev acquired by the neutral component of scalar triplet $\Delta$.

\section{Neutrino Mass}
\label{numass}
Tiny neutrino mass can originate from both type I and type II seesaw mechanisms in our model. As seen from the Yukawa Lagrangian (\ref{yuklag}), the right handed singlet neutrinos acquire a Majorana mass term after the $U(1)_{B-L}$ gauge symmetry gets spontaneously broken by the vev of the singlet scalar field $S$. The resulting type I seesaw formula for light neutrinos is given by the expression,
\begin{equation}
m_{LL}^I=-m_{LR}M_{RR}^{-1}m_{LR}^{T}.
\end{equation}
where $m_{LR} = Y_{\nu} v$ is the Dirac mass term of the neutrinos and $M_{RR} = Y_R\langle S \rangle =Y_R v_{BL}$ is the Majorana mass term of the right handed neutrinos. Demanding the light neutrinos to be of eV scale one needs $M_{RR}$ and hence $v_{BL}$ to be as high as $10^{14}$ GeV without any fine-tuning of dimensionless Yukawa couplings. 

On the other hand, the type II seesaw contribution to neutrino mass comes from the term $f \Delta L L$ in the Yukawa Lagrangian (\ref{yuklag}) if the neutral component of the scalar triplet $\delta^0$ acquires a tiny vev. The scalar triplet can be represented as
\begin{equation}
\Delta =
\left(\begin{array}{cc}
\ \delta^+/\surd 2 & \delta^{++} \\
\ \delta^0 & -\delta^+/\surd 2
\end{array}\right) \nonumber
\end{equation} 
Minimizing the scalar potential gives the approximate value of $v_L$ as
\begin{equation}
v_L = \frac{\lambda_3 v^2 \langle S \rangle}{m^2_{\Delta}}=\frac{\lambda_3 v^2 v_{BL}}{m^2_{\Delta}}
\label{vevvl}
\end{equation}
where $\langle S \rangle = v_{BL}$ is the vev acquired by the singlet scalar field $S$ responsible for breaking $U(1)_{B-L}$ gauge symmetry spontaneously at high scale. Demanding the light neutrinos to be of eV scale one needs $m^2_{\Delta}/v_{BL}$ to be as high as $10^{14}$ GeV without any fine-tuning of dimensionless couplings.

\section{Relic Abundance of Dark Matter}
\label{darkmatter}
The relic abundance of a dark matter particle $\chi$ is given by the the Boltzmann equation
\begin{equation}
\frac{dn_{\chi}}{dt}+3Hn_{\chi} = -\langle \sigma v \rangle (n^2_{\chi} -(n^{eqb}_{\chi})^2)
\end{equation}
where $n_{\chi}$ is the number density of the dark matter particle $\chi$ and $n^{eqb}_{\chi}$ is the number density when $\chi$ was in thermal equilibrium. $H$ is the Hubble expansion rate of the Universe and $ \langle \sigma v \rangle $ is the thermally averaged annihilation cross section of the dark matter particle $\chi$. In terms of partial wave expansion $ \langle \sigma v \rangle = a +b v^2$. Numerical solution of the Boltzmann equation above gives \cite{Kolb:1990vq}
\begin{equation}
\Omega_{\chi} h^2 \approx \frac{1.04 \times 10^9 x_F}{M_{Pl} \sqrt{g_*} (a+3b/x_F)}
\end{equation}
where $x_F = m_{\chi}/T_F$, $T_F$ is the freeze-out temperature, $g_*$ is the number of relativistic degrees of freedom at the time of freeze-out. Dark matter particles with electroweak scale mass and couplings freeze out at temperatures approximately in the range $x_F \approx 20-30$. More generally, $x_F$ can be calculated from the relation 
\begin{equation}
x_F = \ln \frac{0.038g_{eff}m_{PL}m_{DM}<\sigma_{eff} v>}{g_*^{1/2}x_f^{1/2}}
\label{xf}
\end{equation}
The expression for relic density again simplifies to \cite{Jungman:1995df}
\begin{equation}
\Omega_{\chi} h^2 \approx \frac{3 \times 10^{-27} \text{cm}^3 \text{s}^{-1}}{\langle \sigma v \rangle}
\label{eq:relic}
\end{equation}
The thermal averaged annihilation cross section $\langle \sigma v \rangle$ is given by \cite{Gondolo:1990dk}
\begin{equation}
\langle \sigma v \rangle = \frac{1}{8m^4T K^2_2(m/T)} \int^{\infty}_{4m^2}\sigma (s-4m^2)\surd{s}K_1(\surd{s}/T) ds
\end{equation}
where $K_i$'s are modified Bessel functions of order $i$, $m$ is the mass of Dark Matter particle and $T$ is the temperature.

Here we consider the neutral component of the scalar doublet $\phi$ as the dark matter candidate which is similar to the inert doublet model of dark matter discussed extensively in the literature \cite{Ma:2006km,Barbieri:2006dq,Majumdar:2006nt,LopezHonorez:2006gr,ictp,borahcline}. We consider the lighter mass window for the scalar doublet dark matter $m_{DM} \leq M_W$, the W boson mass. Beyond the W boson mass threshold, the annihilation channel of scalar doublet dark matter into $W^+W^-$ pairs opens up reducing the relic abundance of dark matter below observed range for dark matter mass all the way upto around $500$ GeV. We note however, that there exists a region of parameter space $M_W < m_{DM} <160$ GeV which satisfy relic density bound if certain cancellations occur between several annihilation diagrams \cite{honorez1}. For the sake of simplicity, we stick to the low mass region $10 \; \text{GeV} < m_{DM} < M_W$ in this work. We also note that there are two neutral components in the doublet $\phi$, the lighter of which is stable and hence the dark matter candidate. If the mass difference between these neutral scalars $\Delta m = m_{A_0} -m_{H_0}$ is large compared to the freeze-out temperature $T_F$, then the next to lightest neutral scalar play no significant role in determination of dark matter relic density. However, if $\Delta m$ is of the order of freeze-out temperature then $A_0$ can be thermally produced and hence the coannihilations between $H_0$ and $A_0$ during the epoch of dark matter thermal annihilation can play a non-trivial role in determining the relic abundance of dark matter. The annihilation cross section of dark matter in such a case gets additional contributions from coannihilation between dark matter and next to lightest neutral component of scalar doublet $\phi$. This type of coannihilation effects on dark matter relic abundance were studied by several authors in \cite{Griest:1990kh, coann_others}. Here we follow the analysis of \cite{Griest:1990kh} to calculate the effective annihilation cross section in such a case. The effective cross section can given as 
\begin{align}
\sigma_{eff} &= \sum_{i,j}^{N}\langle \sigma_{ij} v\rangle r_ir_j \nonumber \\
&= \sum_{i,j}^{N}\langle \sigma_{ij}v\rangle \frac{g_ig_j}{g^2_{eff}}(1+\Delta_i)^{3/2}(1+\Delta_j)^{3/2}e^{\big(-x_F(\Delta_i + \Delta_j)\big)} \nonumber \\
\end{align}
where, $x_F = \frac{m_{DM}}{T}$ and $\Delta_i = \frac{m_i-m_{DM}}{m_{DM}}$  and 
\begin{align}
g_{eff} &= \sum_{i=1}^{N}g_i(1+\Delta_i)^{3/2}e^{-x_F\Delta_i}
\end{align}
The thermally averaged cross section can be written as
\begin{align}
\langle \sigma_{ij} v \rangle &= \frac{x_F}{8m^2_im^2_jm_{DM}K_2((m_i/m_{DM})x_F)K_2((m_j/m_{DM})x_F)} \times \nonumber \\
& \int^{\infty}_{(m_i+m_j)^2}ds \sigma_{ij}(s-2(m_i^2+m_j^2)) \sqrt{s}K_1(\sqrt{s}x_F/m_{DM}) \nonumber \\
\label{eq:thcs}
\end{align}

\begin{figure}[!h] 
\centering

\includegraphics{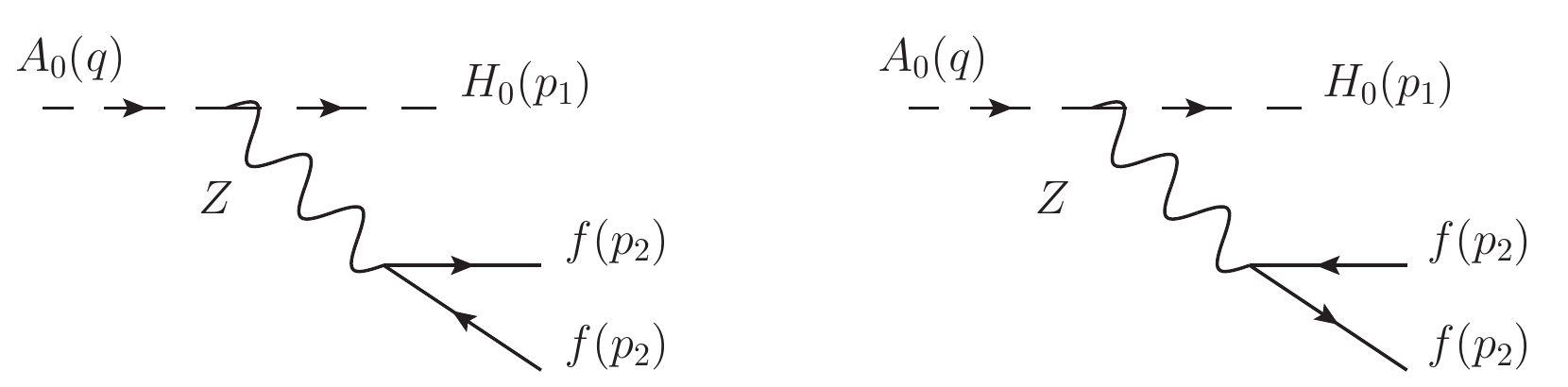}

\caption[]{{\small Decay of $A_0$ into $H_0$ and two fermions (where the conventions are followed from \cite{Dreiner:2008tw} )}}
\label{fig:decay1}
\end{figure}

\begin{figure}[!h] 
\centering
\includegraphics{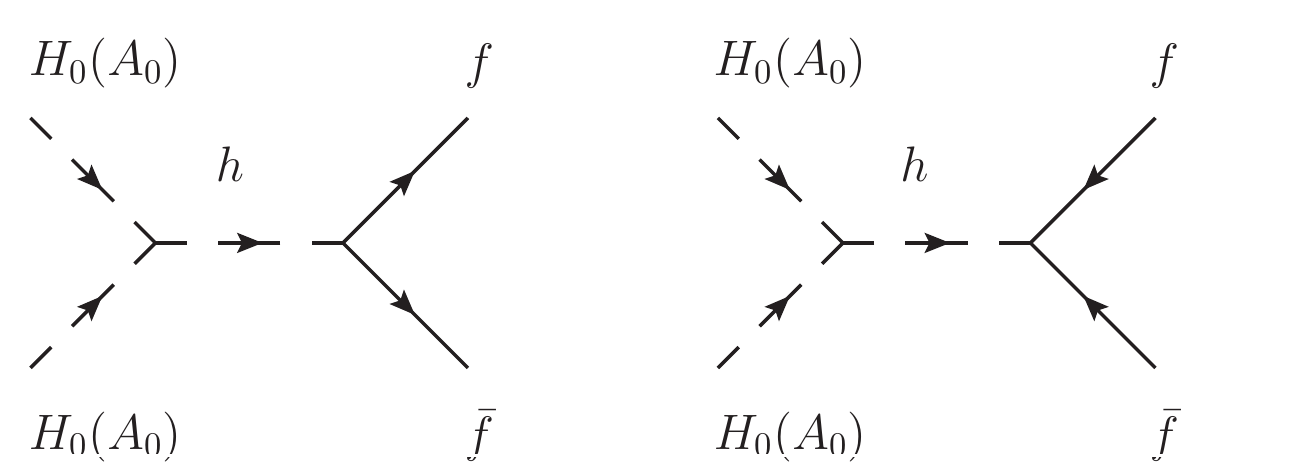}
\caption[]{{\small Self annihilation of $H_0(A_0)$ into two fermions (where the conventions are followed from \cite{Dreiner:2008tw} )}}
\label{fig:feyn1}
\end{figure}

In our model, the lightest neutral component of the scalar doublet $\phi$ is the dark matter candidate. We denote it as $H_0$ and the other neutral component is denoted as $A_0$. Since $A_0$ is heavier than $H_0$, it can always decay into $H_0$ and standard model particles (as shown in figure \ref{fig:decay1}) depending on the mass difference. If the mass difference between $A_0$ and $H_0$ is small enough for $A_0$ to be thermally produced during the epoch of freeze-out then we have to compute both annihilation and coannihilation cross sections to determine the relic abundance. In the low mass regime $(m_{DM} < M_W)$, the self annihilation of either $H_0$ or $A_0$ into SM particles occur through standard model Higgs boson as shown in figure \ref{fig:feyn1}. The corresponding annihilation cross section is given by 
\begin{align}
\sigma_{xx} &= \frac{|Y_f|^2|\lambda_x|^2}{16\pi s}
\frac{\left(s-4m^2_f\right)^{3/2}}{\sqrt{s-4m^2_x}(s-m^2_h + m^2_h\Gamma^2_h)^2}
\label{eq:crossH_0H_0}
\end{align}
where $x\rightarrow H_0,A_0$, $\lambda_x$ is the coupling of $x$ with SM Higgs boson $h$ and $\lambda_f$ is the Yukawa coupling of fermions. $\Gamma_h = 4.15$ MeV is the SM Higgs decay width.

\begin{figure}[!h] 
\centering
\includegraphics[width=1.0\textwidth]{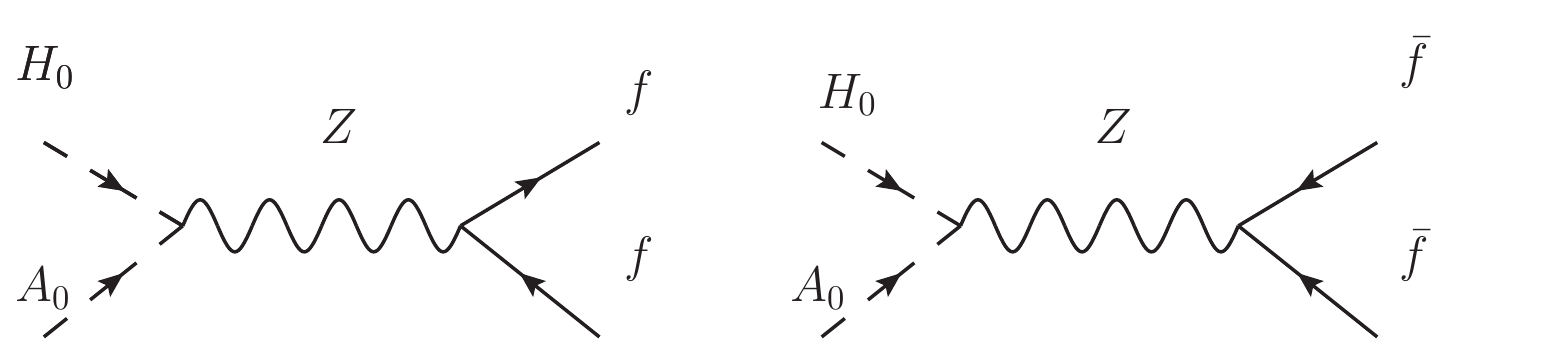}
\caption[]{{\small Coannihilations of $H_0$ and $A_0$ into two fermions (where the conventions are followed from \cite{Dreiner:2008tw} ) }}
\label{fig:feyn}
\end{figure}

The coannihilation of $H_0$ and $A_0$ into SM particles can occur through a $Z$ boson exchange as shown in figure \ref{fig:feyn}. The corresponding cross section is found to be
\begin{align}
\sigma_{H_0A_0} &=  \frac{1}{64 \pi^2 s}\sqrt{\frac{\left(s^2 - 4 m^{2}_fs\right)}{s^2-2(m^2_{H_0}+m^{2}_{A_0})s+(m^{2}_{H_0}-m^{2}_{A_0})^2}} \times \nonumber \\
& \frac{1}{4}\frac{g^4}{c^{4}_W}\frac{1}{\left[(s-m^{2}_z)^2 + m^{2}_z\Gamma^{2}_z\right]}\left[(a^{2}_f+b^{2}_f)\left((m^{2}_{H_0}-m^{2}_{A_0})^2 \right. \right.\nonumber \\ 
&- \left. \left. \frac{1}{3s}\left(s^2 - 2s(m^{2}_{H_0}+ m^{2}_{A_0}) + (m^{2}_{H_0}-m^{2}_{A_0})^2\right) (s-4m^{2}_f)c^{2}_\theta + (s-2m^{2}_f)(s-(m^2_{H_0}+m^{2}_{A_0}))\right ) \right. \nonumber \\
&+ \left. a_fb_fm^{2}_f(s-(m^{2}_{H_0} + m^{2}_{A_0}))\right]
\label{eq:crossH_0A_0}
\end{align}
where $a_f = T^{f}_3 - s^{2}_WQ_f; \quad b_f = -s^{2}_WQ_f$. $\Gamma_z = 2.49$ GeV is the Z boson decay width.

We use these cross sections to compute the thermal averaged annihilation cross section given in equation (\ref{eq:thcs}). Instead of assuming a particular value of $x_F$, we first numerically find out the value of $x_F$ which satisfies the following equation
\begin{align}
e^{x_F} - \ln \frac{0.038g_{eff}m_{PL}m_{DM}<\sigma_{eff} v>}{g_*^{1/2}x_F^{1/2}} &= 0
\end{align}
which is nothing but a simplified form of equation (\ref{xf}). For a particular pair of $\lambda_{DM}$ and $m_{DM}$, we use this value of $x_F$ and compute the thermal averaged cross section$<\sigma_{eff} v>$ to be used for calculating relic abundance using equation (\ref{eq:relic}).

We also calculate the lifetime of $A_0$ to make sure that $A_0$ is not long lived enough to play a role of dark matter in the present Universe. The decay width of $A_0$ is given by
\begin{align}
\Gamma_{A_0} &= \int_{s_2}\int_{s_{3-}}^{s_{3+}} \frac{1}{32(2\pi)^3 m^3_{A_0} }f(s_2,s_3)ds_2 ds_3 \nonumber \\
f(s_2,s_3)&= \frac{N_cg^4}{4c^4_W}\bigg[(a_f^2+b_f^2)\big[(m^2_{A_0}-m^2_{H_0}-2m^2_f)^2-(s_2-s_3)^2-4a_fb_fm^2_f(m^2_{A_0}+m^2_{H_0}+s_3+s_2)\big]\bigg]\nonumber \\
&\times \frac{1}{\bigg[(4m^2_fm^2_{A_0}m^2_{H_0}-m^2_Z-s_3-s_2)^2-m^2_Z\Gamma^2_Z\bigg]} 
\end{align}
where,
\begin{align}
s_{3\pm} &= m^2_f + m^2_{H_0} \frac{1}{s_2}\bigg[(m^2_{A_0}-s_2-m^2_f)(s_2-m^2_f+m^2_{H_0})\pm \lambda^{1/2}(s_2,m^2_{A_0},m^2_f)\lambda^{1/2}(s_2,m^2_f,m^2_{H_0})\bigg]\nonumber \\
s_2 &\in \bigg[(m_{H_0} + m_f)^2,(m_{A_0} - m_f)^2\bigg] 
\end{align}
Now if the difference $\Delta m = m_{A_0} - m_{H_0} $= 50 keV then $m_{A_0}$ will decay into $H_0$ and neutrinos only and its lifetime is 
$\Gamma_{A_0}^{-1}\hbar = 1.67962\times10^6$ s. If the difference is 5 MeV then $A_0$ can decay into up-quarks, electrons and neutrinos such that $\Gamma_{total} = \Gamma_u + \Gamma_e + \Gamma_{\nu}$ and therefore the lifetime will be 
$\Gamma_{total}^{-1}\hbar = 8.3659$ s. But if the difference is around 1 GeV then $A_0$ can decay into strange-quarks, down-quarks, up-quarks, muons, electrons and neutrinos such that $\Gamma_{total} = \Gamma_s + \Gamma_d + \Gamma_u+ \Gamma_{\mu} + \Gamma_e + \Gamma_{\nu}$ and therefore the lifetime will be 
$\Gamma_{total}^{-1}\hbar = 1.2 \times 10^{-11}$ s. WIMP dark matter typically freeze-out at temperature $T_F \sim m_{DM}/x_F$ where $x_f \sim 20-30$. This can roughly be taken to be the time corresponding to the electroweak scale $t_{EW} \sim 10^{-11}$ s. Thus, for all mass differences under consideration the lifetime of $A_0$ falls much below the present age of the Universe. Hence, the present dark matter relic density is totally contributed by the abundance of the lightest stable neutral scalar $H_0$.


\begin{figure} 
\centering
\begin{tabular}{c}
\epsfig{file=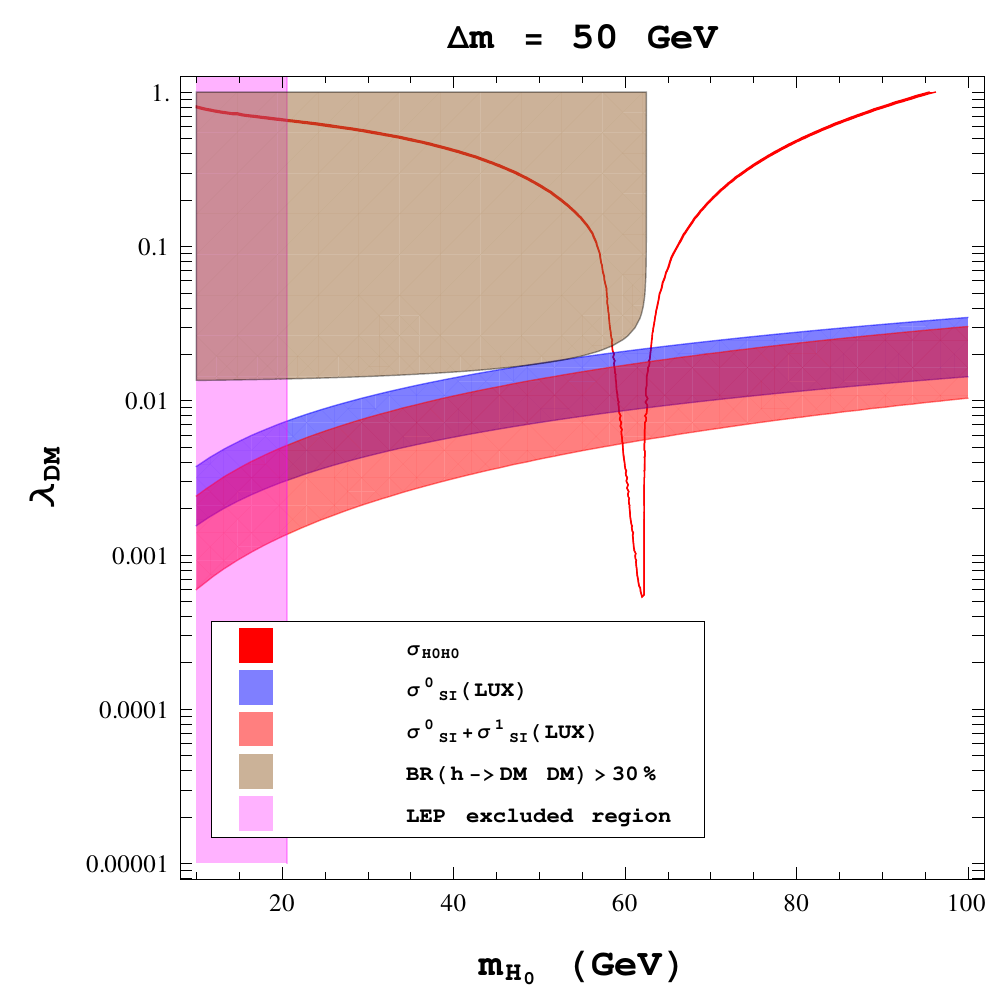,width=1.0\linewidth,clip=}\\
\end{tabular}

\caption{Plot of $\lambda_{DM}$ versus $m_{H_0}$ satisfying dark matter relic density constraint: The red colored v shaped region is for dark matter $H_0$ without any coannihilation, for mass difference of $50$ GeV between $A_0$ and $H_0$. The blue and red shaded regions are the direct detection exclusion limits from LUX experiment incorporating tree level and one-loop level scatterings respectively, the brown shaded region is the excluded region from SM Higgs invisible decay width constraint $BR (h \rightarrow \text{DM}\;\text{DM}) < 30\%$. The pink shaded region is the region forbidden by the LEP I precision measurement of $Z$ boson decay width.}
\label{fig:p20}
\end{figure}

\begin{figure} 
\centering
\begin{tabular}{c}
\epsfig{file=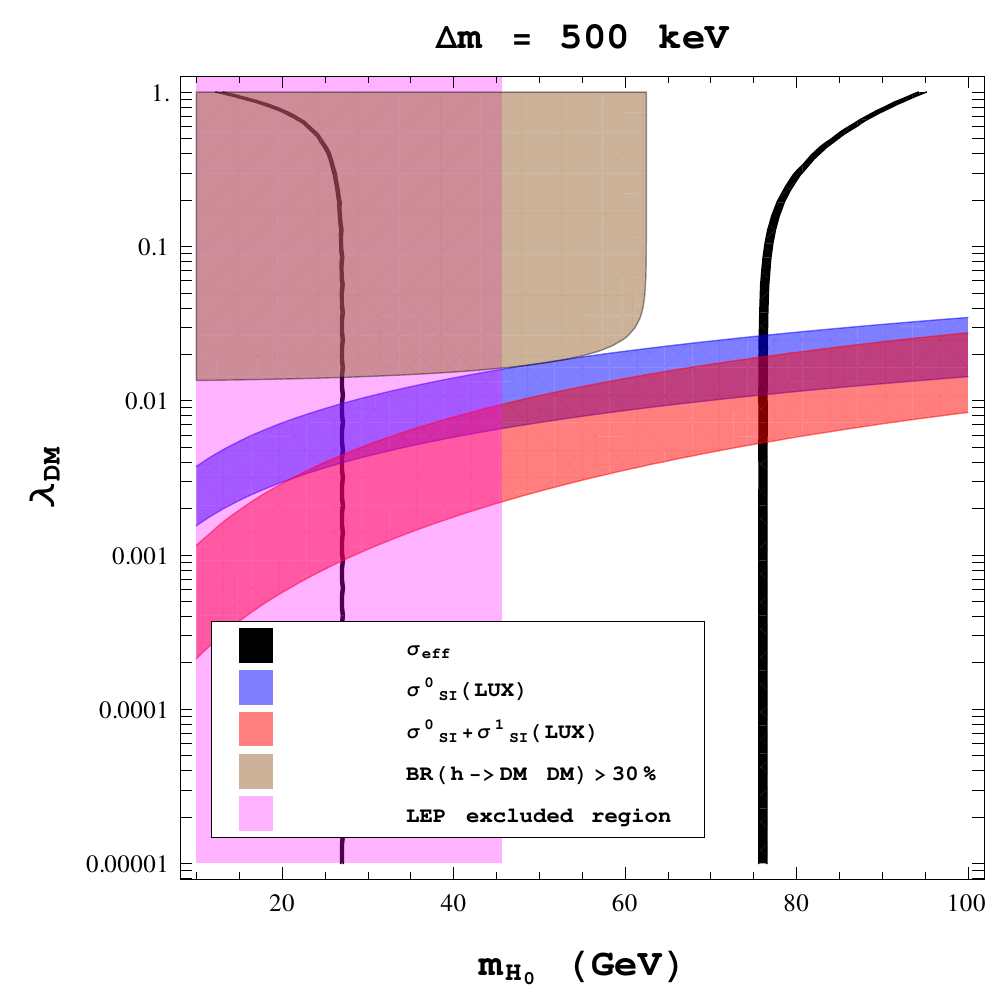,width=1.0\linewidth,clip=}\\
\end{tabular}

\caption{Plot of $\lambda_{DM}$ versus $m_{H_0}$ satisfying dark matter relic density constraint: the black colored region is for dark matter $H_0$ incorporating coannihilation with mass difference of $500$ keV between $A_0$ and $H_0$. The blue and red shaded regions are the direct detection exclusion limits from LUX experiment incorporating tree level and one-loop level scatterings respectively, the brown shaded region is the excluded region from SM Higgs invisible decay width constraint $BR (h \rightarrow \text{DM}\;\text{DM}) < 30\%$. The pink shaded region is the region forbidden by the LEP I precision measurement of $Z$ boson decay width.}
\label{fig:p21}
\end{figure}

\section{Results}
\label{results}
We follow the approach and use the expressions discussed in the previous section to calculate the dark matter relic density. We first calculate the relic density of dark matter $H_0$ without considering coannihilation. We use the constraint on dark matter relic density (\ref{dm_relic}) and show the allowed parameter space in terms of $\lambda_{DM} = \lambda_1 + \lambda_2$, the dark matter-SM Higgs coupling and $m_{H_0}$, the dark matter mass. The results are shown as the red v-shaped region in figure \ref{fig:p20}. We then allow coannihilation between $H_0$ and $A_0$ and show the allowed parameter space in the same $\lambda_{DM}-m_{H_0}$ plane. This corresponds to the black region in figure \ref{fig:p21}. The plot \ref{fig:p21} corresponds to mass splitting between $A_0$ and $H_0$: $\Delta m = 500$ keV.

In addition to the Planck 2013 constraints on dark matter relic density (\ref{dm_relic}), there is also a strict 
limit on the spin-independent dark matter-nucleon cross section coming from
direct detection experiments, most recently from LUX experiment\cite{LUX}. The
relevant scattering cross section in our model is given
by \cite{Barbieri:2006dq}
\begin{equation}
 \sigma_{SI} = \frac{\lambda^2_{DM}f^2}{4\pi}\frac{\mu^2 m^2_n}{m^4_h m^2_{DM}}
\label{sigma_dd}
\end{equation}
where $\mu = m_n m_{DM}/(m_n+m_{DM})$ is the DM-nucleon reduced mass. A recent estimate of the Higgs-nucleon coupling $f$ gives $f = 0.32$ \cite{Giedt:2009mr} although the full range of allowed values is $f=0.26-0.63$ \cite{mambrini}. We take the minimum upper limit on the dark matter-nucleon spin independent cross section from LUX experiment \cite{LUX} which is $7.6 \times 10^{-46} \; \text{cm}^2$ and show the exclusion line in figure \ref{fig:p20}, \ref{fig:p21} with a label $\sigma^0_{SI}$. The exclusion line gets broaden as shown in blue in figure \ref{fig:p20}, \ref{fig:p21} due to the uncertainty factor in Higgs-nucleon coupling. It should be noted that the spin independent scattering cross section written above (\ref{sigma_dd}) is only at tree level. Since dark matter in this model is part of a doublet under $SU(2)_L$, one-loop box diagrams involving two $W$ bosons or two $Z$ bosons can also give rise to direct detection cross section. As discussed in details by authors of \cite{oneloop}, in the low mass regime of inert doublet dark matter, this one-loop correction is maximum in the resonance region $m_{DM} \sim m_h/2$. The ratio $\sigma_{SI} (\text{one-loop})/ \sigma_{SI} (\text{tree-level})$ is approximately $100$ in this region. We calculate the scattering cross section corresponding to one-loop diagrams mentioned in \cite{oneloop}. The final result not only depends upon $\lambda_{DM}$ but also on the charged Higgs mass. For simplicity, we take $m_{H^{\pm}} = m_{A_0}$. The one-loop result is shown as the exclusion line in figure \ref{fig:p20}, \ref{fig:p21} with label $\sigma^0_{SI}+\sigma^1_{SI}$. Thus, in the absence of co-annihilation only a small region of the parameter space near the resonance is left from direct detection bound. In the presence of co-annihilation, more regions of parameter space gets allowed as seen from figure \ref{fig:p21}.
\begin{figure}
\centering
\includegraphics[width=1.0\textwidth]{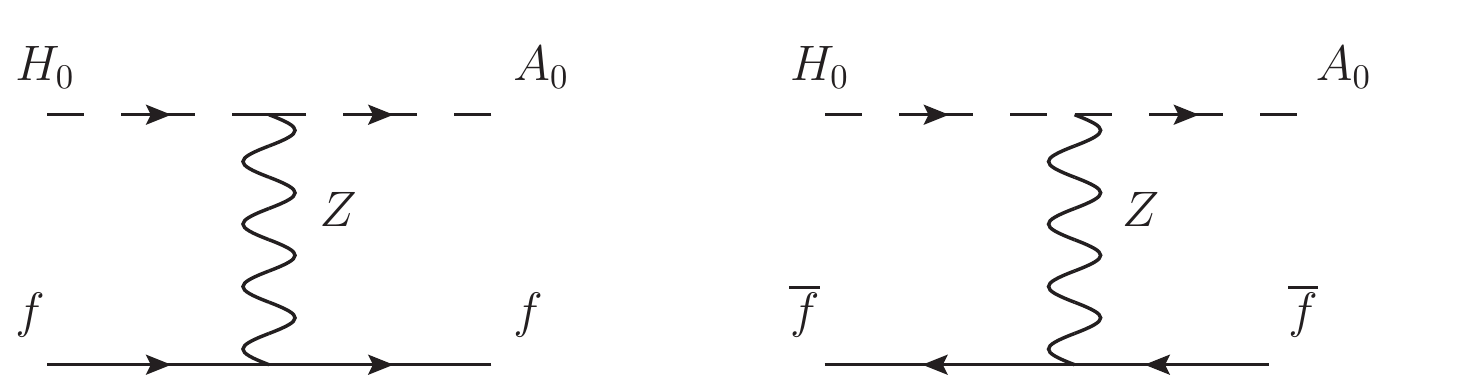}
\caption[]{{\small Forward scattering of $H_0 f(\overline{f})\rightarrow A_0f(\overline{f})$ through t-channel $Z$ boson (where the conventions are followed from \cite{Dreiner:2008tw} )}}
\label{fig:inelastic}
\end{figure}

Apart from the SM Higgs mediated scattering, there can be one more DM-nucleon scattering cross section due to the same interaction giving rise to coannihilation between $H_0$ and $A_0$. This corresponds to DM scattering off nuclei into $A_0$ through a Z boson exchange giving rise to an inelastic DM-nucleon scattering as shown in figure \ref{fig:inelastic}. The cross-section for such a process can be calculated as
     \begin{align}
    |\mathcal{M}_{H_0f\rightarrow A_0f}|^2 &= \frac{g^2}{4\cos^2 \theta_W}\frac{1}{(t-m^2_Z)^2-m^2_Z\Gamma^2_Z}\bigg{[}2(a^2_f+b^2_f)\bigg{[}\frac{1}{4}
    (m^2_{H_0}-m^2_{A_0}+s-u)^2\nonumber \\
      &+\frac{1}{2}(t-2m^2_f)(2(m^2_{H_0}m^2_{A_0}-t)\bigg{]}+4a_fb_fm^2_f(2(m^2_{A_0}+m^2_{H_0})-t)\bigg{]} 
	  \nonumber \\
	  \frac{d\sigma}{d\Omega_{cm}} &= \frac{1}{64\pi^2s}\sqrt{\frac{s^2-2(m^2_{A_0}+m^2_f)s+(m^2_{A_0}-
	    m^2_f)^2}{s^2-2(m^2_{H_0}+m^2_f)s+(m^2_{H_0}-
	      m^2_f)^2}}|\mathcal{M}_{H_0f\rightarrow A_0f}|^2
	  \label{eq_cross_t}
	  \end{align}

Due to the strong Z boson coupling to DM in our model, such a scattering can give rise to an inelastic cross section which faces severe limits from direct detection experiments like Xenon100 \cite{Xenoninelastic}. Such inelastic dark matter within inert doublet dark matter was studied by authors in \cite{idmARINA}. They show that such inelastic dark matter scenario in inert doublet model is consistent with exclusion limits from direct detection experiments only when dark matter relic abundance is below the observed abundance. However, such a scattering process is kinematically forbidden if the mass difference between $H_0$ and $A_0$ is more than the kinetic energy of dark matter $H_0$. Taking the typical speed of WIMP dark matter to be $v \approx 270 \; \text{km/s} \sim 10^{-3}c$, the kinetic energy of a $100$ GeV WIMP is around $50$ keV. Thus for mass differences more than $50$ keV, dark matter relic abundance can get affected by $A_0$, but the direct detection cross section remain unaffected.

We also impose collider bounds by noting that the precision measurement of the $Z$ boson decay width at LEP I forbids the $Z$ boson decay channel $Z \rightarrow H_0 A_0$ which requires $m_{H_0}+m_{A_0} > m_Z \Rightarrow 2m_{H_0} > m_Z -\Delta m$. The excluded region $m_{H_0} < (m_Z-\Delta m)/2$ is shown as a pink shaded region in figure \ref{fig:p21}. Apart from LEP I constraint on $Z$ decay width, LEP II constraints also rule out models satisfying $m_{H_0} < 80$ GeV, $m_{A_0} < 100$ GeV and $m_{A_0}-m_{H_0} > 8$ GeV \cite{Lundstrom:2008ai}. As we see from figure \ref{fig:p21}, the allowed region of $\lambda_{DM}-m_{DM}$ parameter space including coannihilation satisfy these constraints as the mass difference is not more than 8 GeV.

We further impose the constraint that invisible decay of the SM Higgs boson $h \rightarrow H_0 H_0$ do not dominate its decay width. Recent measurement of the SM Higgs properties constrain the invisible decay width to be below $30\%$ \cite{HiggsInv}. The invisible decay width is given by
\begin{equation}
\Gamma_{\rm inv} = {\lambda^2_{DM} v^2\over 64 \pi m_h} 
\sqrt{1-4\,m^2_{DM}/m^2_h}
\end{equation}
We show the parameter space ruled out by this constraint on invisible decay width as the brown shaded region for $m_{H_0} < m_h/2$ in figure \ref{fig:p21}. It can be easily seen that this imposes a weaker constraint than the DM direct detection constraint from LUX experiment.

In our analysis we have taken the mass differences between $A_0$ and $H_0$ to be $\Delta m = 500$ keV. As noted in section \ref{model}, the mass difference between $A_0$ and $H_0$ is given by $\Delta m^2 = 4\mu_{\phi \Delta} v_L$. Thus, for $\Delta m = 500$ keV and dominant type II seesaw such that $v_L = 0.1$ eV, the trilinear mass term $\mu_{\phi \Delta} \sim 600$ GeV. However, if we keep the trilinear mass term fixed at say, the $U(1)_{B-L}$ symmetry breaking scale, then different mass differences $\Delta m$ will correspond to different strengths of type II seesaw term. If we fix the trilinear mass term $\mu_{\phi \Delta}$ to be $10^9$ GeV say, then $\Delta m = 500 \; \text{keV}, 5\; \text{MeV}, 1 \; \text{GeV}$ will correspond to $v_L = 10^{-16}, 10^{-14}, 10^{-9} \; \text{GeV}$ respectively. The first two examples correspond to a case of sub-dominant type II seesaw similar to the ones discussed in \cite{typeI+II}, whereas the third example $\Delta m = 1$ GeV corresponds to a type II dominant seesaw. Comparing this with equation (\ref{vevvl}), one gets 
$$ \frac{\lambda_3 v_{BL}}{m^2_{\Delta}} = 10^{-20}, 10^{-18}, 10^{-13} \; \text{GeV}^{-1}$$ respectively. This can be achieved by suitable adjustment of the symmetry breaking scales, the bare mass terms in the Lagrangian as well as the dimensionless couplings.

We note that, in conventional inert doublet dark matter model, this mass squared difference is $\lambda_{IDM} v^2$ where $\lambda_{IDM}$ is a dimensionless coupling. If we equate this mass difference to a few hundred keV, then the dimensionless coupling $\lambda_{IDM}$ has to be fine tuned to $10^{-12}-10^{-10}$. Such a fine tuning can be avoided in our model by suitably fixing the symmetry breaking scales and the bare mass terms in the Lagrangian.
\section{Conclusion}
\label{conclude}
We have studied an abelian extension of SM with a $U(1)_{B-L}$ gauge symmetry. The model allows the existence of both type I and type II seesaw contributions to tiny neutrino masses. It also allows a naturally stable cold dark matter candidate: lightest neutral component of a scalar doublet $\phi$. Type II seesaw term is generated by the vev of a scalar triplet $\Delta$. We show that, in our model the vev of the scalar triplet not only decides the strength of type II seesaw term, but also the mass splitting between the neutral components of the scalar doublet $\phi$. If the vev is large (of the order of GeV, say), then mass splitting is large and hence the next to lightest neutral component $A_0$ plays no role in determining the relic abundance of $H_0$. However, if the vev is small such that the mass splitting is below $5-10\%$ of $m_{H_0}$, then $A_0$ can play a role in determining the relic abundance of $H_0$. In such a case, dark matter $(H_0)$ relic abundance gets affected due to coannihilation between these two neutral scalars. We compute the relic abundance of dark matter in both the cases: without and with coannihilation and show the change in parameter space. We incorporate the latest constraint on dark matter relic abundance from Planck 2013 data and show the allowed parameter space in the $\lambda_{DM}-m_{DM}$ plane, where $\lambda_{DM}$ is the dark matter SM Higgs coupling and $m_{DM}$ is the mass of dark matter $H_0$. We show the parameter space for mass splitting $\Delta m = 500 \; \text{keV}$. We point out that unlike the conventional inert doublet dark matter model, here we do not have to fine tune dimensionless couplings too much to get such small mass splittings between $A_0$ and $H_0$. This mass splitting can be naturally explained by the suitable adjustment of symmetry breaking scales and bare mass terms of the Lagrangian. It is interesting to note that for sub-dominant type II seesaw case, the dark matter relic abundance gets affected by coannihilation whereas for dominant type II seesaw case, usual dark matter relic abundance calculation applies taking into account of self-annihilations only. 

We also take into account the constraint coming from dark matter direct detection experiments like LUX experiment on spin independent dark matter nucleon scattering. We incorporate the LEP I bound on $Z$ boson decay width which rules out the region $m_{H_0}+m_{A_0} < m_Z$. We then incorporate the constraint on invisible SM Higgs decay branching ratio from measurements done at LHC experiment. We show that after taking all these relevant constraints into account, there still remains viable parameter space which can account for dark matter as well as neutrino mass simultaneously. Thus our model not only gives rise to a natural dark matter candidate, but also provides a natural way to connect the dark matter relic abundance with the neutrino mass term from type II seesaw mechanism.

\begin{acknowledgments}
AD likes to thank Council of Scientific and Industrial Research, Govt. of India for financial support through Senior Research Fellowship
(EMR No. 09/466(0125)/2010-EMR-I).
\end{acknowledgments}

\bibliographystyle{apsrev}

\end{document}